\documentclass[11pt,twoside]{article}

\usepackage{asp2006}
\usepackage{epsf}
\usepackage{lscape}
\usepackage{graphicx}

\def    \bJ     {\bf J}
\def    \bB     {\bf B}

\begin{document}
\title{Alignment of Dust by Radiative Torque: Recent Developments}   
\author{A. Lazarian and Thiem Hoang}   
\affil{Department of Astronomy, University of Wisconsin-Madison}    

\begin{abstract} 
Alignment of dust by radiative torques (RATs) has proven to be the most promising mechanism to explain alignment in various astrophysical environments, from comet atmospheres to circumstellar accretion disks, molecular clouds, and diffuse interstellar gas. Recent years have been marked by intensive attempts to provide proper theoretical treatment of the alignment process. We discuss some of the major advances, which include, first of all, formulating of the analytical model of RATs. This model was shown to reproduce well the torques acting on actual irregular dust grains and allowed studies of the parameter space for  which the alignment happens with long axes perpendicular and parallel to the magnetic field. Such a study resulted in an important conclusion that, without any paramagnetic relaxation, the RAT alignment always happens for interstellar grains with long axes perpendicular to the magnetic field. Incidentally, this conclusion is not true for the alignment of large (i.e. $a> 10^{-4}$~cm) grains that are present, e.g. in accretion disks, as these grains can demonstrate both the alignment with long axes parallel and perpendicular to the magnetic field, although the alignment with long grain axes perpendicular to the magnetic field is preferential in many cases, which can also be identified with the analytical model. Additional recent advances include the description of effects of gaseous bombardment and pinwheel torques on grains aligned by RATs. Very counterintuitively, the gaseous bombardment was shown in some cases to increase the degree of alignment by knocking out grains from the positions of imperfect alignment when the grains rotate slowly to more stable positions of perfect alignment where grains rotate fast. In terms of pinwheel torques, important revisions have been made in the Lazarian \& Draine model of grain flipping and thermal trapping. Those, however, do not change the major conclusion that very small grains (i.e. $a<3\times 10^{6}$~cm) should be marginally aligned. Recent work made the RAT alignment a predictive theory which is ready for quantitative modeling of astrophysical polarization. The right timing of this work is not only due to renewed efforts in terms of astrophysical polarimetry, but also due to the interest to the polarized foreground arising from the attempts to measure CMB polarization. In particular, we predict that the microwave emission from the Zodiacal dust presents an important contaminant, which should be included into foreground polarization templates. 
\end{abstract}


\section{Grain Alignment: History of ideas}   

Grain alignment has a reputation of being a very tough astrophysical problem of a very long standing.
Indeed, for a long time since the discovery of dust-induced starlight extinction polarization in 1949 (Hall 1949; Hiltner 1949) the mechanism of alignment was both enigmatic and illusive.  Works by great minds like Lyman Spitzer and Edward Purcell moved the field forward in terms of understanding of grain dynamics, but, nevertheless, left observations and theory poorly connected.

An extended discussion of the history of the grain alignment subject can be found in the review by Lazarian (2003). Below we briefly mention several milestones in our understanding of grain alignment. The first extended review of grain alignment theory is by Purcell (1969). However, the subject is so rapidly developing that even the most recent extended review by Lazarian (2007) is outdated in a number of aspects.    

Several mechanisms were proposed and elaborated to various degree (see Lazarian 2007 for a review), including the "textbook solution", namely, 
the paramagnetic Davis-Greenstein (1951) mechanism, which matured through intensive work since its introduction (e.g. Jones \& Spitzer 1967, Purcell 1979, Spitzer \& McGlynn 1979, Mathis 1986, Roberge et al. 1993, Lazarian 1997, Roberge \& Lazarian 1999). The mechanical stochastic alignment was pioneered by Gold (1951), who concluded that supersonic flows should align grains rotating thermally. Further advancement of the mechanical alignment mechanism (e.g. Lazarian 1994, 1995a) allowed one to extend the range of applicability of the mechanism, but left it as an auxiliary process, nevertheless. Mechanical regular alignment of helical grains discussed in Lazarian (1995b), Lazarian, Goodman, Myers (1997), Lazarian \& Hoang (2007ab) seems to be more promising as it can align grains within subsonic flows. In any case, currently the mechanism based on radiative torques looks as the most promising (see Andersson \& Potter 2007, Lazarian 2007, Whittet et al. 2008). Thus we focus our present contribution on this mechanism. 

The effect of alignment induced by radiative torques was discovered by  Dolginov \& Mytrophanov (1976). They considered a grain which exhibited a difference in the cross-section for right-handed and left-handed photons. They noticed that scattering of unpolarized light by such a grain resulted in its spin-up. However, they could not quantify the effect and therefore their pioneering work was mostly neglected for the next 20 years\footnote{The initial work made predictions for
functional dependences of RATs, which were not confirmed by further research (Lazarian \& Hoang 2007, Hoang \& Lazarian 2008c). It has additional problems, as some of the processes, e.g. internal relaxation (Purcell 1979) and thermal wobbling arising from coupling of rotational and vibrational degrees of freedom (Lazarian 1994) had not been a part of accepted grain dynamics by 1976. Nevertheless, as we discuss in more details in Hoang \& Lazarian (2008c), the ground-breaking nature of this study should not be questioned.}. The explosion of interest to the radiative torques we owe to Bruce Draine, who realized that the torques can be treated with the modified version of the discrete dipole approximation code, namely, DDSCAT, by Draine \& Flatau (1994). Empirical studies in Draine (1996), Draine \& Weingartner (1996, 1997), Weingartner \& Draine (2003) demonstrated that the magnitude of torques is substantial for irregular shapes studied. After that it became impossible to ignore the radiative torque alignment. Later, the spin-up of grains by radiative torques was demonstrated in laboratory conditions (Abbas et al. 2004). 

It should be stressed, however, that the reliable predictions for the alignment degree cannot be obtained with the "brute force approach". Indeed, radiative torques depend on many parameters, e.g. grain shape, grain size, radiation wavelength, grain composition, the angle between the radiation direction and the magnetic field. It is not practical to do numerical calculations for this vast multidimensional parameter space. Therefore the important empirical studies above had limited predictive powers and were used to demonstrated the radiative torque effects sometimes using one grain shape, one grain size, one wavelength of light, and one direction of the light beam with respect to the magnetic field. The quantitative stage of radiative torque studies required theoretical models describing radiative torques. In Lazarian \& Hoang (2007a) we proposed a simple model of radiative torques which allowed a good analytical description of the radiative torque alignment. This model was elaborated and extended in Lazarian \& Hoang (2008) and Hoang \& Lazarian (2008abc). Further on, we shall abbreviate RAdiative Torques to RATs.

\section{Analytical Model for Radiative Torques} 

\begin{figure}
\includegraphics[width=0.4\textwidth,angle=270]{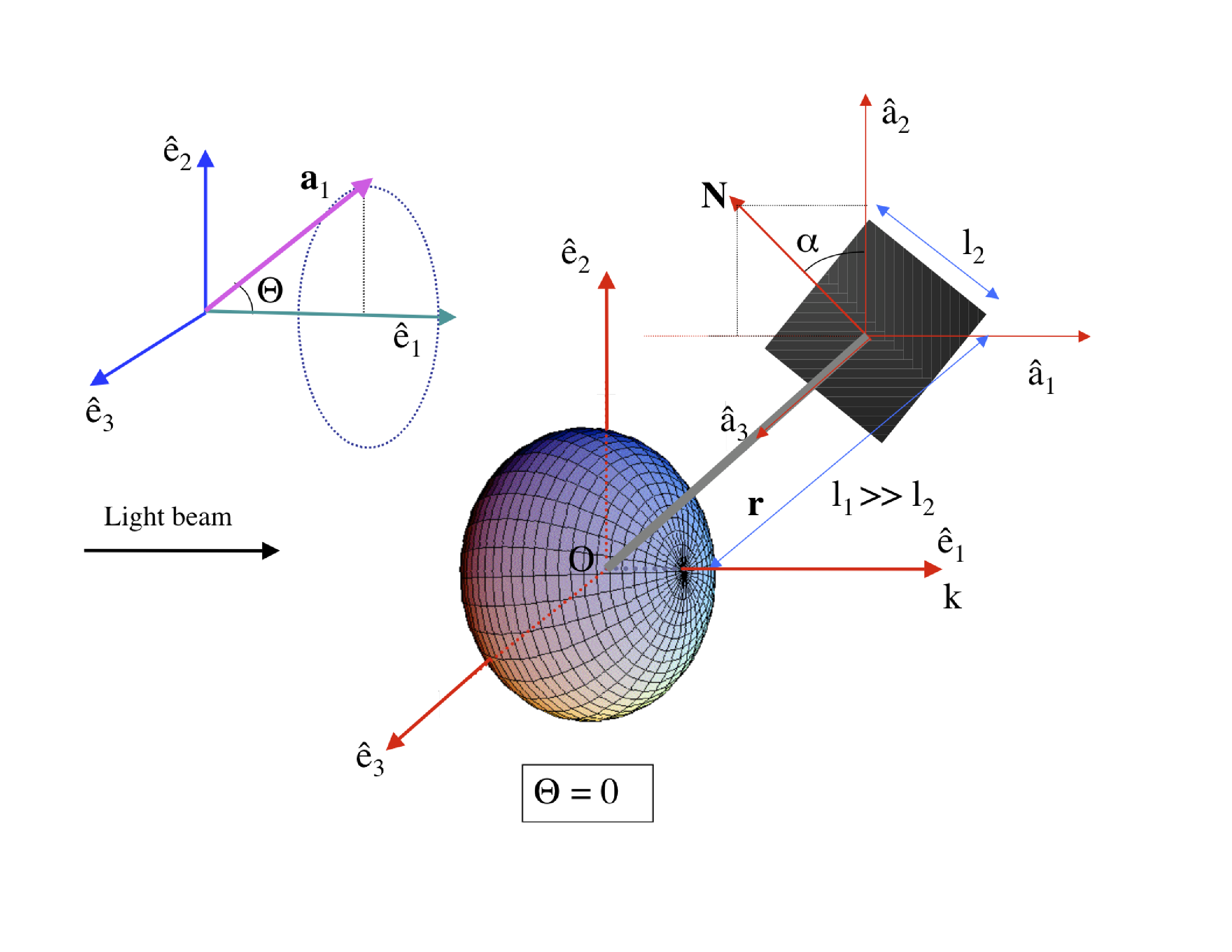}
\includegraphics[width=0.35\textwidth,angle=270]{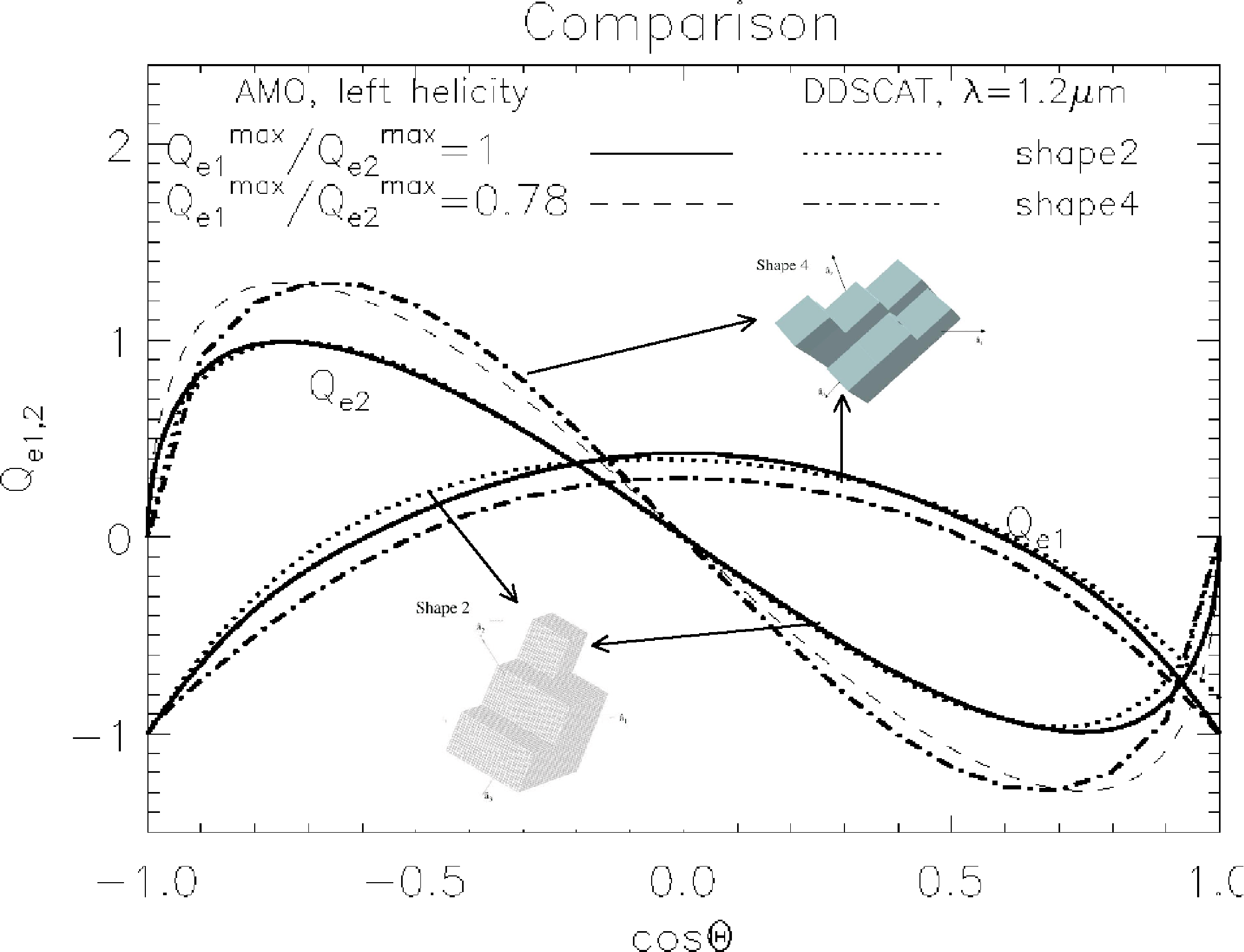}
\caption{
\small
{\it (a) Left panel}.-- A model of a ``helical'' grain,
that consists of a spheroidal body with a mirror at an angle $\alpha$ attached to it ($\alpha$ is chosen to be $\pi/4$ in the standard LH07 model).
 The ``scattering coordinate system'' which
illustrates the definition of torque components: ${\bf a}_1$ is directed
along the maximal inertia axis of the grain; ${\bf k}$ is the direction of radiation.
The projections of normalized radiative torques $Q_{e1}$,
$Q_{e2}$ and $Q_{e3}$ are calculated in this reference frame.  {\it (b) Right panel}.-- The model reproduces well the radiative torques as it clear from the comparison of the functional form of torques predicted by the model with the RATs obtained via DDSCAT calculations for irregular grains. Images of irregular
grains, shapes 2 and 4 are also shown. From Lazarian \& Hoang 2007a.}
\label{AMO}
\end{figure}

Lazarian \& Hoang (2007a, henceforth LH07) showed that a simple model in Fig.~\ref{AMO} reproduces well the essential basic properties of RATs. The model consists of an ellipsoidal grain with a mirror attached to
its side. LH07 termed the model AMO, which is an abbreviation from Analytical MOdel. Note,  that grains can be both  
``left-handed'' and  ``right-handed''. For our grain model to become ``right handed'' the mirror should be turned  by 90 degrees. 
Our studies with DDSCAT confirmed that actual irregular grains also vary in handedness. A substantial difference in the RATs acting on
right and left handed irregular grains was a source of earlier confusion. 
The torques obtained analytically in the assumption of geometric optics for the model in Fig.~\ref{AMO} were shown to be in good agreement with the torques calculated numerically with DDSCAT for irregular grains. To make a proper comparison LH07 chose a system of reference with the direction of light along a vector ${\bf e_1}$, and the grain axis of maximal inertia moment ${\bf a}_1$ being in the ${\bf e_1}$, ${\bf e_2}$ plane. As a result, we showed that for the problem of alignment only torques $Q_{e1}$ and $Q_{e2}$ mattered. The third component $Q_{e3}$ happens to induce grain precession only, which for most situations is subdominant to the Larmor precession of the grain in the interstellar magnetic field (see more discussion in LH07). Interestingly enough, the conclusion of $Q_{e3}$ is not important in terms of the RAT alignment is also true for the presence of thermal fluctuation (see Hoang \& Lazarian 2008a) and inefficient internal relaxation (see Hoang \& Lazarian 2008c) when the alignment of angular momentum and axis ${\bf a}_1$ is not enforced\footnote{It was shown in LH07 that the only component of RATs present for an ellipsoidal grain is $Q_{e3}$. Naturally, this component cannot produce the RAT alignment, as the helicity of an ellipsoidal grain is zero.}. 

\begin{figure}
\includegraphics[width=0.5\textwidth]{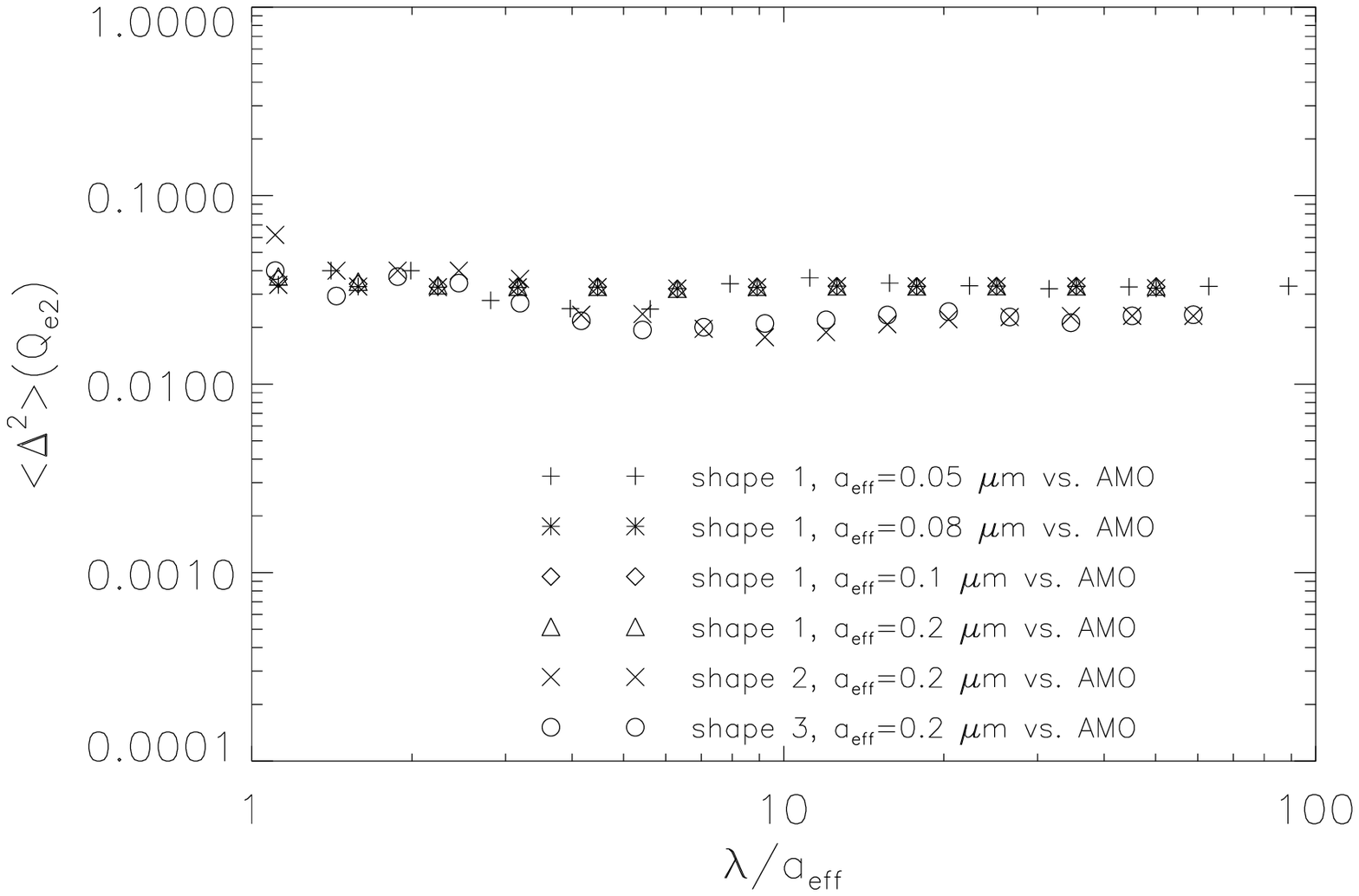}
\includegraphics[width=0.5\textwidth]{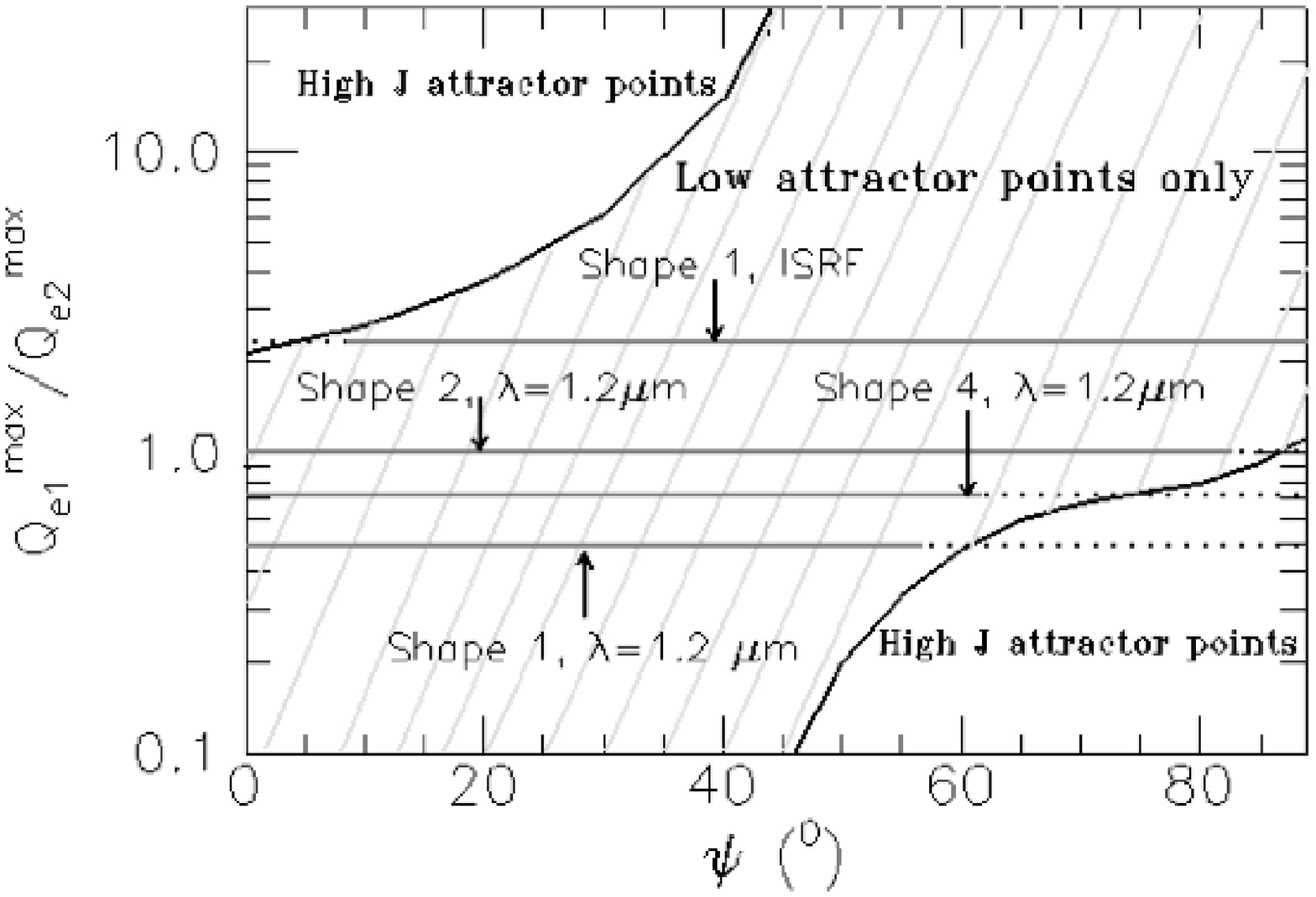}
\caption{\small {\it (a) Left panel}: Numerical comparison of the torques calculated with DDSCAT for irregular grains for different wavelength and the analytical model (AMO) of a helical grain. The quantity
$\Delta^2$ is defined by Eq.~(\ref{chi_eq}). {\it (b) Right panel}: Parameter space for which grains have only low-$J$ attractor point and both low-$J$ and high attractor point. $\psi$ is the angle between the direction towards a point radiation source and magnetic field. In the situation when the high-$J$ attractor point
is present grains eventually get there and demonstrate perfect alignment. In the situation when only low-$J$ attractor point is present, the alignment is partial. From Lazarian \& Hoang 2007a. }
\label{chi}
\end{figure}

The functional dependences of torques $Q_{e1}(\Theta)$ and $Q_{e2}(\Theta)$, where $\Theta$ is an angle between the axis ${\bf a}_{1}$ and the radiation direction, were shown to be very similar for the analytical model in Fig.~\ref{AMO} and irregular grains subject to radiation of different wavelengths. 
In Figure~\ref{AMO} this correspondence is shown for two irregular grains (Shape 2 and Shape 4 in LH07) and AMO. This remarkable correspondence is illustrated in Fig.~\ref{chi}a using a function:

\begin{equation}
\langle \Delta^{2}\rangle(Q_{e2})=\frac{1}{\pi (Q_{e2}^{max})^{2}} \int^{\pi}_{0} [Q_{e2}^{irregular}(\Theta) -Q_{e2}^{model} (\Theta)]^{2} d\Theta,
\label{chi_eq}
\end{equation}
which characterizes the deviation of the torques $Q_{e2}$ calculated numerically for irregular grains from the analytical prediction in the LH07 model. 

Observing Figure~\ref{AMO}a it is easy to notice that while the functional dependence of torque components $Q_{e1}(\Theta)$ and $Q_{e2}(\Theta)$ coincides for grains of various shapes, their amplitudes vary for different  grains and different radiation wavelengths. In fact, LH07 showed that the radiative torque alignment can be fully determined if  
the ratio $q^{max}=Q_{e1}^{max}/Q_{e2}^{max}$ is known. In terms of practical calculations, this enormously simplifies the calculations
of radiative torques: instead of calculating two {\it functions} $Q_{e1}(\Theta)$ and $Q_{e2}(\Theta)$ it
is enough to calculate just two {\it values} $Q_{e1}^{max}$ and $Q_{e2}^{max}$. According to LH07 the maximal value of the function $Q_{e1}(\Theta)$ is achieved for $\Theta=0$ of the function $Q_{e2}(\Theta)$ is achieved at $\Theta=\pi/4$. In other words, one can use a {\it single number} $q^{max}=Q_{e1}^{max}/Q_{e2}^{max}=Q_{e1}(0)/Q_{e2}(\pi/4)$ instead of
{\it two functions} to characterize grain alignment. Thus, it is possible to claim that the $q^{max}$-ratio is as important for the alignment as the grain axis
ratio for producing polarized radiation by aligned grains.

Studying the RAT alignment LH07 corrected the treatment of grain dynamics in Draine \& Weingartner (1997). As a result, instead of cyclic trajectories in the latter paper, they obtained a situation when, instead of accelerating grains, RATs were slowing grains down. This slowing down in realistic circumstances, e.g.
in the presence of thermal fluctuations (see Hoang \& Lazarian 2008a) did not bring grains to a complete stop, but resulted in creating of low-$J$ attractor point, in agreement with an earlier empirical study in Weingartner \& Draine (2003). Thus, apart from high attractor points, LH07 found that for a range of
$q^{max}$ and the angle $\psi$ between the radiation beam and magnetic field direction only low-$J$ attractor points exist. A later study by Hoang \& Lazarian (2008a) established that when low-$J$ and high-$J$ attractor points coexist, the high-$J$ points are more stable and therefore external stochastic driving, e.g.
arising from gaseous bombardment, brings grains to high-$J$ attractor points. This transfer can take several damping times, but for the steady state interstellar alignment in the situation when a high-$J$ attractor point exists, one can safely assume that grains are aligned with high-$J$.

When does the alignment happens at low-$J$? Figure~\ref{chi}b shows predictions for the existence of low-$J$ and high-$J$ attractor points for the analytical model (AMO) for the parameter space given by $q^{max}$ and the angle $\psi$. Individual horizontal lines correspond to particular grain shapes with a given $q^{max}$. For the interstellar radiation field (ISRF) the calculations of $q^{max}$ are performed using torques averaged over the interstellar spectrum
of wavelengths (see LH07 for more details). We see that the correspondence in terms of predicting the distribution of high-$J$ and low-$J$ attractor points is
also good, which, however, is not surprising due to the good correspondence between the functional dependences obtained for the AMO and irregular grains depicted in Figure~\ref{AMO}a.

\begin{figure}
\includegraphics[width=0.5\textwidth]{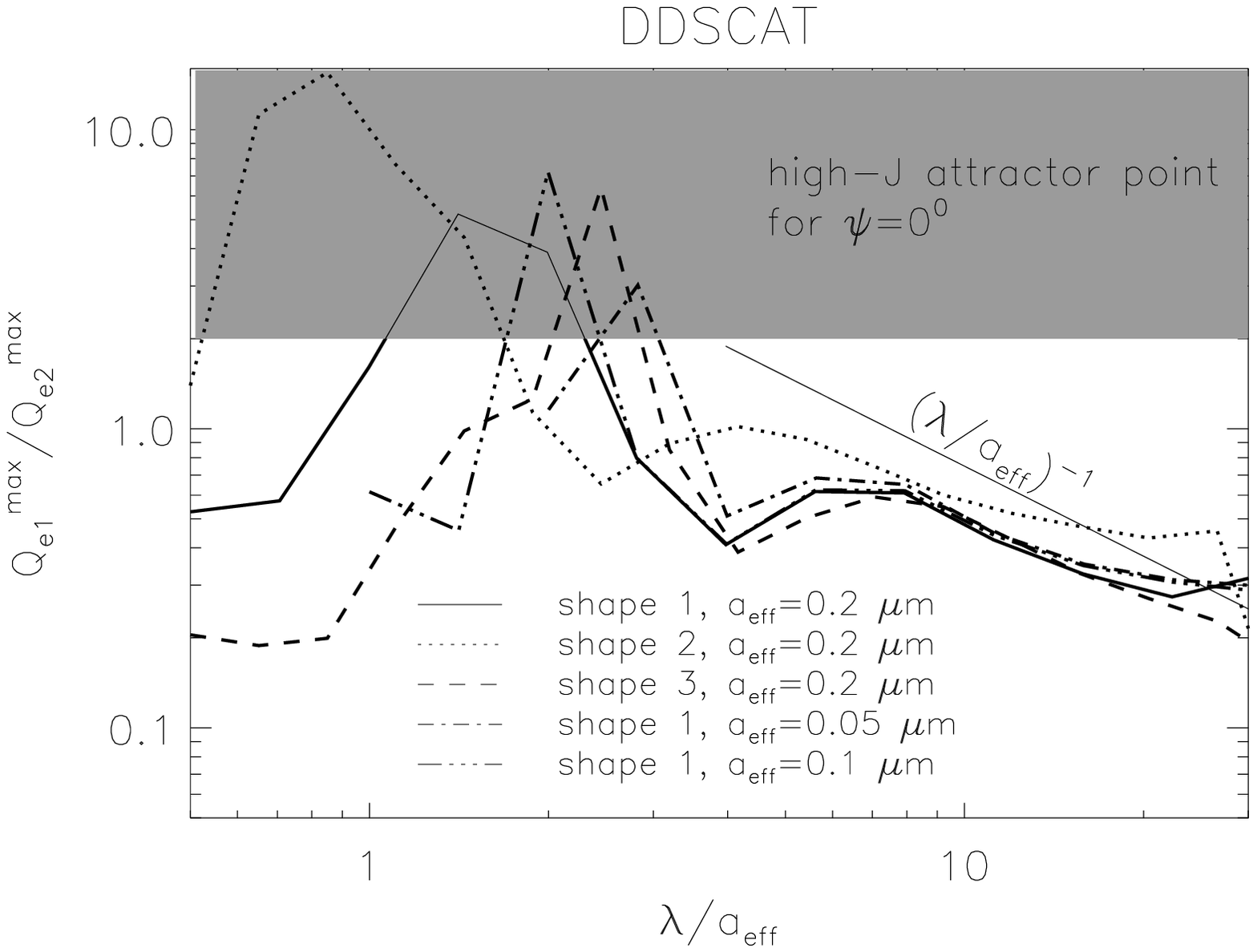}
\includegraphics[width=0.5\textwidth]{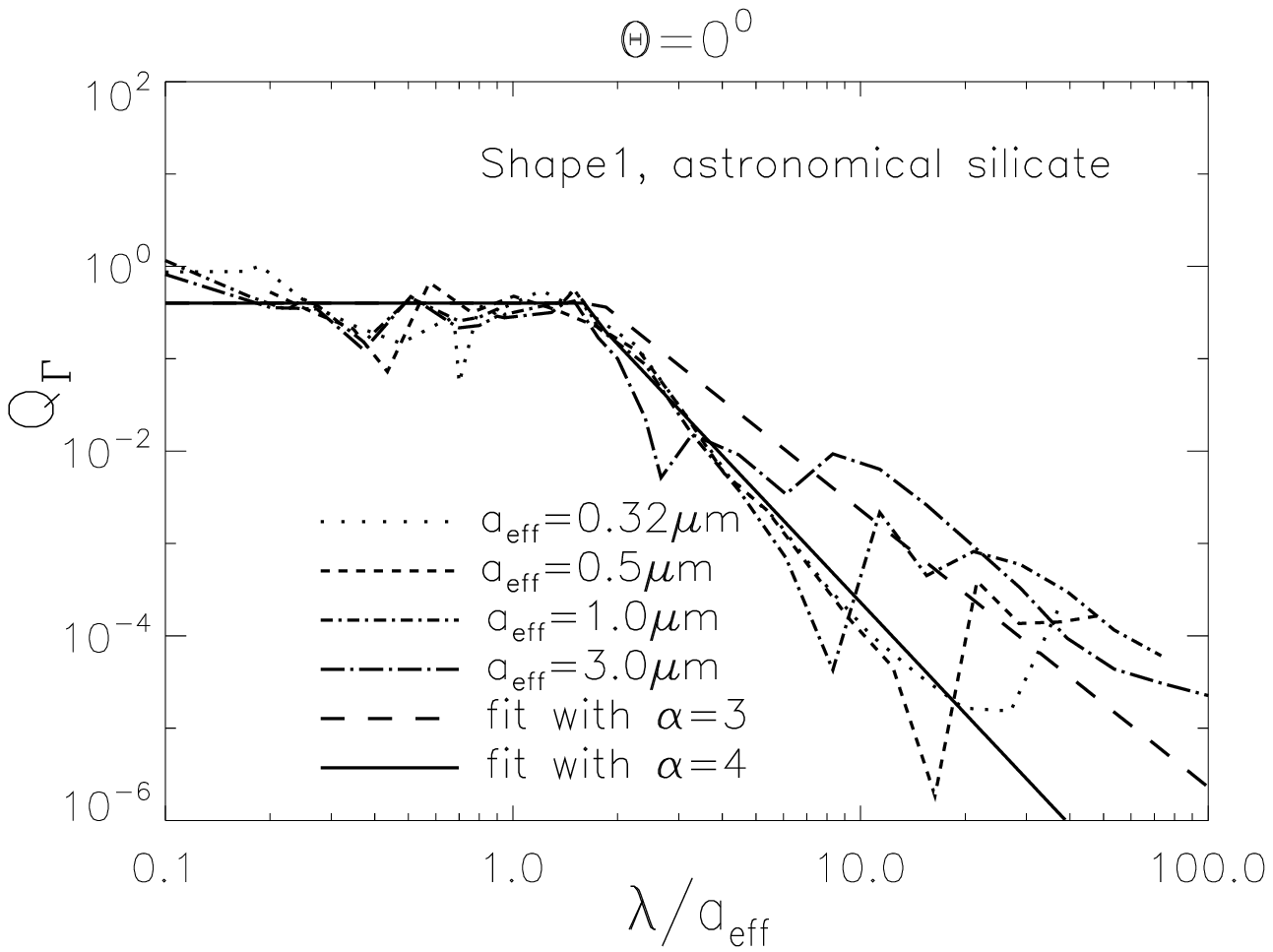}
\caption{\small {\it (a) Left panel}: The magnitude of the ratio $q=Q_{e1}^{max}/Q_{e2}^{max}$ that characterize the radiative torque alignment of grains depends on both grain shape and the wavelength of radiation. {\it (b) Right panel}: Normalized torques for grains of different sizes and wavelengths. The torque 
amplitude is proportional to radiation intensity. The most efficinet alignment is for grains larger than $\lambda/2$. However, the alignment of grains substantially smaller than  the radiation wavelength can also be present provided that the radiation is strong enough. From Lazarian \& Hoang 2007a. }
\label{wavelength}
\end{figure}

The torques $Q_{e1}$ and $Q_{e2}$ vary with the wavelength. This results in the change of $q^{max}$ (see Figure~\ref{wavelength}a), and the magnitude of
the radiative torques. For the latter LH07 presented a simple numerical fit shown in Figure~\ref{wavelength}b. This fit together with the analytical description of RATs in AMO, substantially simplifies modeling of the RAT alignment. 

\section{Why do grains get aligned with long axes perpendicular to ${\bf B}$?}
\label{s:why}

Observations testify that interstellar grains tend to get aligned with long axes perpendicular to the ambient magnetic field, the fact that was frequently used to argue that the Davis-Greenstein (1951) mechanism is responsible for the alignment. Can radiative torques explain this observational fact?
\begin{figure}[h]
\includegraphics[width=0.45\textwidth, angle=270]{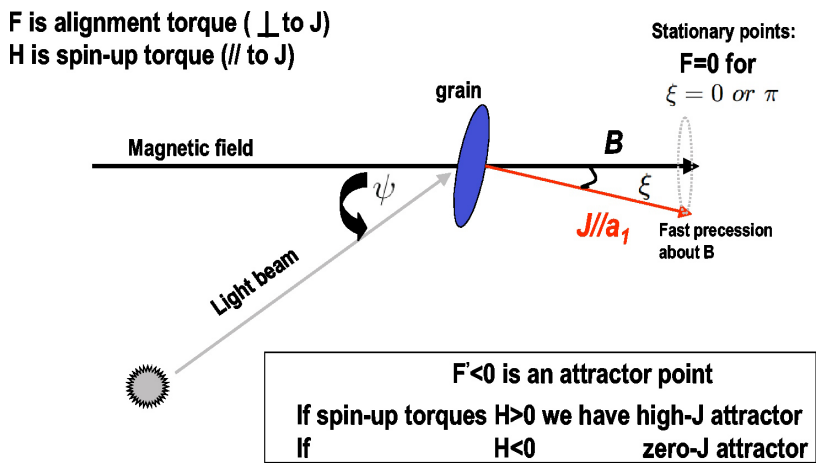}
\caption{\small A simplified explanation of the grain alignment by radiative torques. The grain, which is depicted as a spheroid in the figure, in fact, should be irregular to experience non-zero radiative torque. 
The positions ${\bf J}$ parallel (or anti-parallel) to ${\bf B}$ correspond to the stationary points as at these positions the component of torques that changes the alignment angle vanishes. As 
internal relaxation makes ${\bf J}$ aligned with the axis $a_1$ of the maximal moment of grain inertia, the grain gets aligned with long axes perpendicular to ${\bf B}$.}
\label{alignment}
\end{figure}

Fig.~\ref{alignment} illustrates why radiative torques tend to align grains the ``right way'', i.e. in agreement with observations. Interstellar grains experience internal relaxation that tends to make them rotate about their axis of maximal moment of inertia. Therefore, it is sufficient to follow the dynamics of angular momentum to determine grain axes alignment. Let us call the component of torque parallel to ${\bf J}$ the {\it spin-up torque}
${\bf H}$ and perpendicular to ${\bf J}$ the {\it alignment torque} ${\bf F}$. The angular momentum ${\bf J}$ is precessing about magnetic field due to the magnetic moment of a grain (see
Dolginov \& Mytrophanov 1976). The alignment torques ${\bf F}$ are perpendicular to ${\bf J}$ and therefore as  ${\bf J}$ gets parallel to ${\bf B}$ the fast precession of the grain makes the torques averaged over ${\bf J}$ precession vanish as $\xi\rightarrow 0$. Thus the positions corresponding to ${\bf J}$ aligned with ${\bf B}$ are stationary points, irrespectively of the functional forms of radiative torques, i.e.
of components $Q_{e1}(\Theta)$ and $Q_{e2}(\Theta)$. In other words, grain can stay aligned with $\xi=0$ or $\pi$.

The arguments above are quite general, but they do not address the question whether there are other stationary points, e.g. whether the alignment can also happen with ${\bf J}$ perpendicular to ${\bf B}$.
To answer this question one should use the actual expressions for  $Q_{e1}(\Theta)$ and $Q_{e2}(\Theta)$. The analysis in Lazarian \& Hoang (2007a) shows that there is, indeed, a range of angles
between the direction of radiation and the magnetic field for which grains tend to aligned in a "wrong" way, i.e. with long axes parallel to magnetic field. However, this range of angles is rather narrow and does not exceed several degrees. Moreover, the "wrong" alignment corresponds to the positions for which the spin-up torques ${\bf H}$ are negative, which induces grain alignment with low angular momentum\footnote{Within this short report we do not go into the rather complex details of grain thermal wobbling which stabilize the value of $J$ at some fraction of the thermal value. The corresponding discussion is presented in Lazarian \& Hoang (2007a) and Hoang \& Lazarian (2008a).}. Grain thermal wobbling at low-$J$ attractor point (Lazarian 1994, Lazarian \& Roberge 1997) induces the variations in the angle that is typically exceeds this range. Thus a remarkable fact emerges: grains get always aligned with long axes perpendicular to ${\bf B}$!

\section{Grain dynamics: grains wobbling, flipping and thermally trapped}
\label{wobbling}
Grain dynamics is rather complex and it should be accounted for if measures of alignment are sought. 
\subsection{Internal relaxation}
\label{w1}

 To produce the observed starlight polarization, grains must be aligned, with their
 {\it long axes} perpendicular to the magnetic field. The degree of alignment 
 involves alignment not only of the grain's angular momenta ${\bf J}$ with respect to
 the external magnetic field ${\bf B}$, but also the alignment of the grain's
 long axes with respect to ${\bf J}$.
The latter alignment is frequently termed the "internal alignment".

Purcell (1979, henceforth P79) was the first to consider internal relaxation as the course of
internal alignment. He showed that the rates of internal relaxation of energy in wobbling grains
are much faster than the time of grain alignment. This induced many researchers to think that
all grains rotate about the axis of their maximal inertia moment and the internal alignment is always perfect.
 Lazarian (1994) corrected
this conclusion by showing that the fluctuations associated with the internal dissipation should induce the wobbling, which amplitude increases as the effective rotational temperature of grains approaches that of grain material.

The rate of internal relaxation is an important characteristic of grain dynamics. Spitzer \& McGlynn (1979), Lazarian \& Draine (1997, 1999ab), Lazarian \& Hoang (2008), Hoang \& Lazarian (2008b) demonstrated its crucial significance for various aspects of grain alignment. P79 was the first to evaluate the rates of internal relaxation, taking into account the inelastic effects\footnote{A more recent study of inelastic relaxation is provided in Lazarian \& Efroimsky (1999).} and a particular effect that he discovered himself, the Barnett relaxation. As we know, the Barnett effect is the magnetization of a paramagnetic body as the result of its rotation. Purcell (1979) noticed that the wobbling rotating body would experience changes of the magnetization arising from the changes of the direction of rotation in grain axes. This would entail fast relaxation, with
a characteristic time about a year for a $10^{-5}$~cm grains. Lazarian \& Draine (1999a) identified $10^6$ times stronger relaxation related to nuclear spins and Lazarian \& Hoang (2008) showed that superparamagnetic grains, e.g. grains with  magnetic inclusions (see Jones \& Spitzer 1967, Mathis 1986, Martin 1994), exhibit enhanced internal relaxation. 

\begin{figure}
\includegraphics[width=0.5\textwidth]{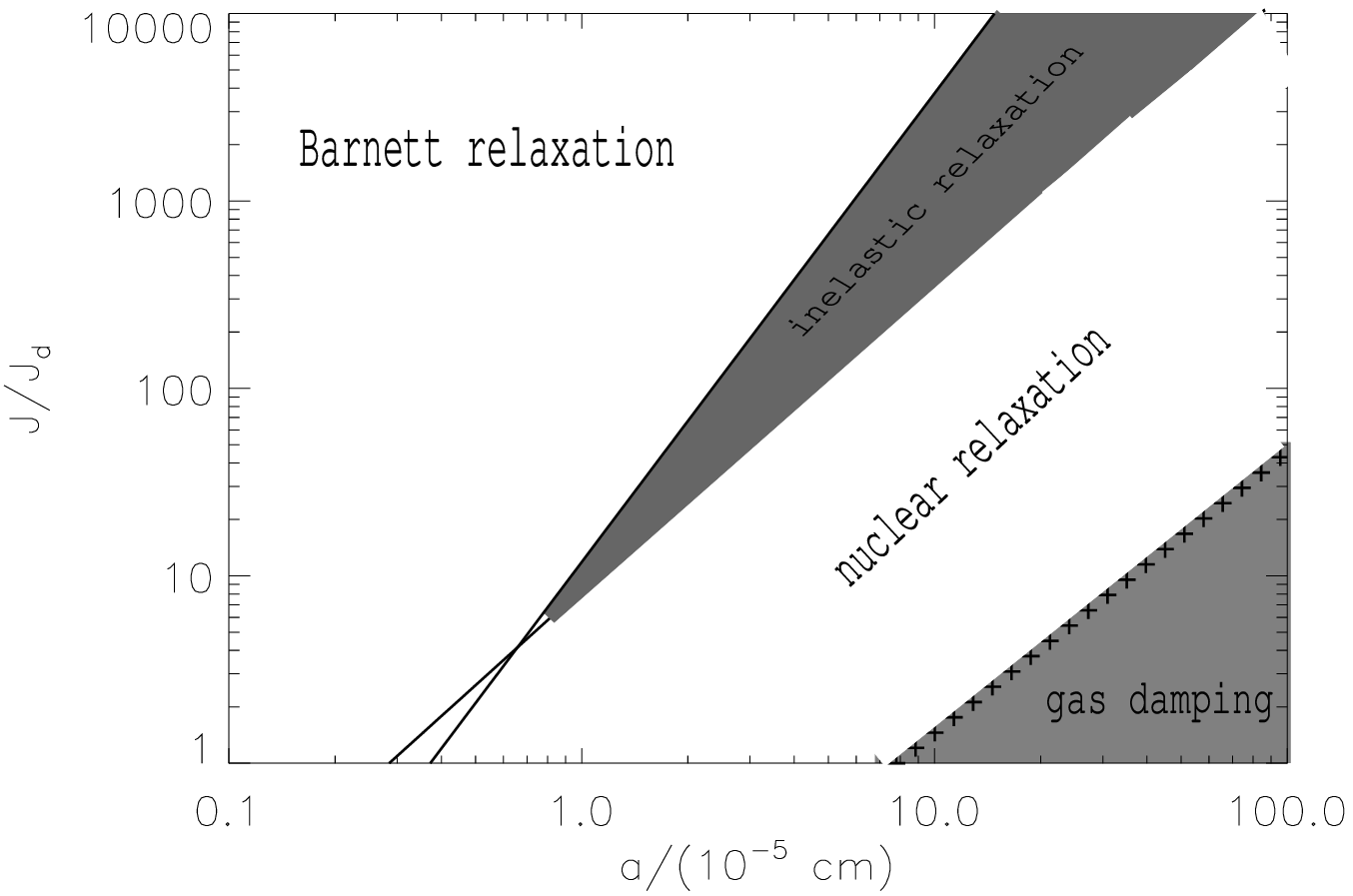}
\includegraphics[width=0.4\textwidth]{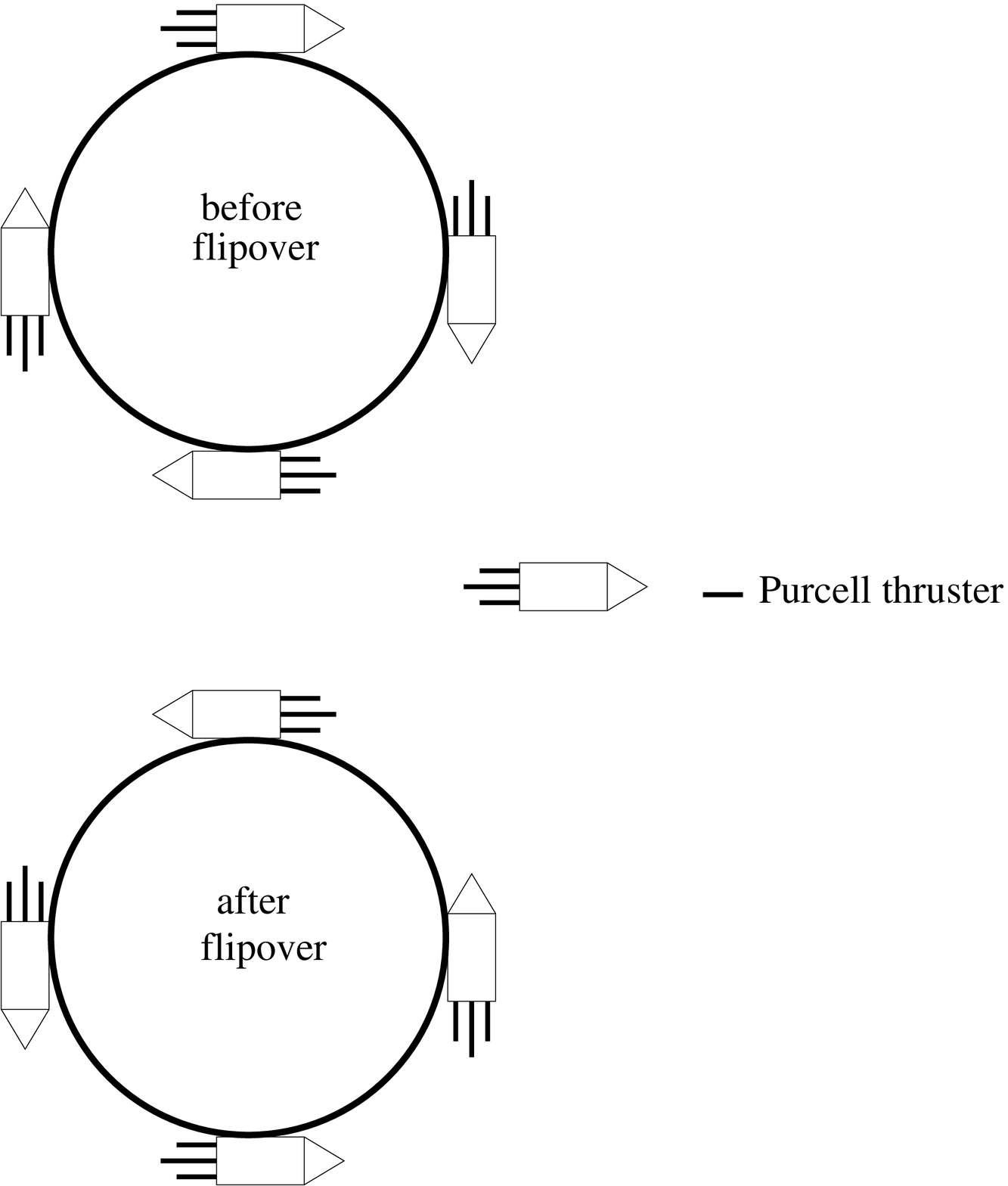}
\caption{\small {\it (a) Left Panel}--The parameter space for the dominance  of the Barnett, nuclear and inelastic relaxation for grains of different sizes in terms of grain angular momentum normalized by its value corresponding to the rotation at thermal rate for temperature equal to the dust temperature. Rates of Barnett relaxation are from Purcell (1979), nuclear relaxation are from Lazarian \& Draine (1999b) and for inelastic relaxation are from Lazarian \& Efroimsky (1999). {\it (b) Right Panel}-- Illustration of pinwheel torques and their averaging to zero as a result of flipping. 
 A grain acted upon by pinwheel
torques before and after a flipover event. As the grain flips, the direction
of torques alters. In the picture the H$_2$ formation sites are depicted as thrusters.}
\label{relax}
\end{figure}

The relative role of internal relaxation for the Barnett, nuclear and inelastic relaxation is shown in Fig.~\ref{relax}a. The calculations of inelastic relaxation from Lazarian \& Efroimsky (1999) are used for the plot. 
These rates are important for grain dynamics. In particular, they influence how grain crossovers happen (see \S\ref{regular}). 
\subsection{Regular pinwheel torques}
\label{regular}    

 P79 realized that
grains may rotate at a rate much faster that the thermal rate if they are subject to systematic torques. P79
 identified three separate systematic torque
mechanisms: inelastic scattering of impinging atoms when gas and grain
temperatures differ, photoelectric emission, and H$_2$ formation on grain
surfaces (see Fig.~\ref{relax}b). Below we shall refer to the latter as "Purcell's
torques". These were shown to dominate the other two for typical conditions in
the diffuse ISM (P79). The existence of systematic H$_2$ torques is expected
due to the random distribution over the grain surface of catalytic sites of
H$_2$ formation, since each active site acts as a minute thruster emitting
newly-formed H$_2$ molecules. A later study of uncompensated torques in
Hoang \& Lazarian (2008b) added additional systematic torques to the list, namely,
torques arising from plasma-grain interactions and torques arising from emission
of radiation by an irregular grains. Radiative torques arising from the interaction
of the {\it isotropic} radiation with an irregular grain (Draine \& Weingartner 1996) 
also represent the systematic torques fixed in the grain body. We shall call all the systematic
torques above {\it pinwheel torques} to distinguish them from RATS, which are systematic torques arising 
from anisotropic flow of photons or atoms
interacting with an helical grain (see below).

\begin{figure}
\includegraphics[width=0.3\textwidth]{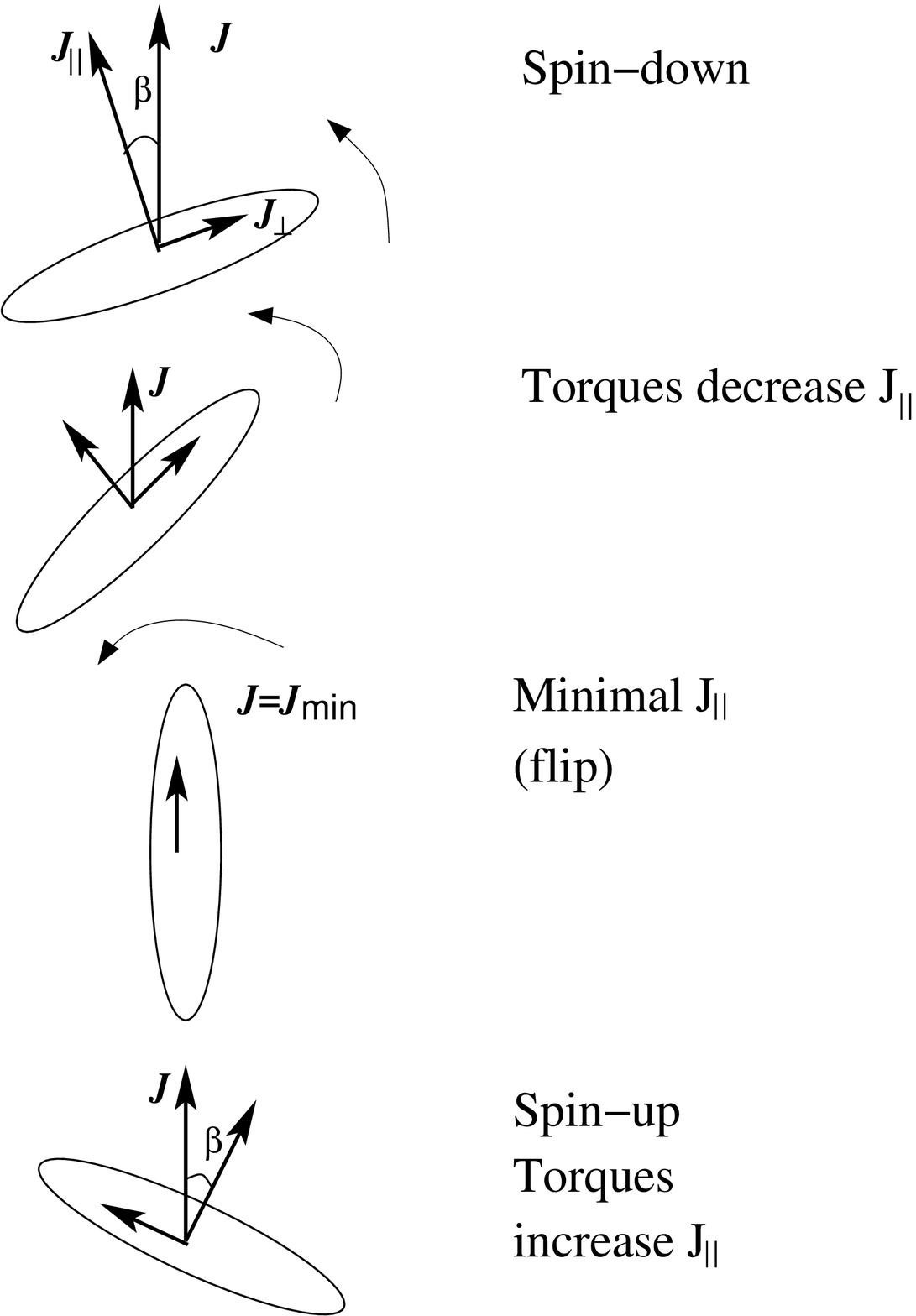}
\includegraphics[width=0.5\textwidth]{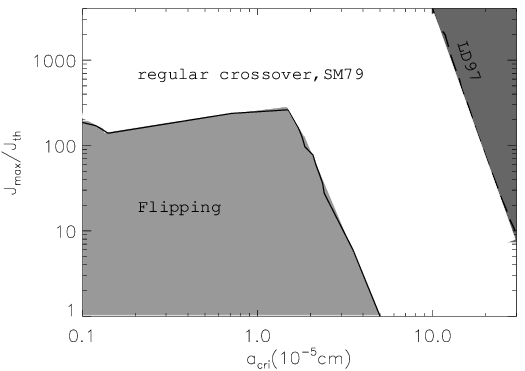}
\caption{ \small  {\it (a) Left panel}-- A regular crossover event as described by Spitzer \& McGlynn (1979).
The systematic torques nullify the amplitude of the ${\bf J}$ component
parallel to the axis of maximal inertia, while preserving the other component,
$J_{\bot}$. If $J_{\bot}$ is small then the grain is susceptible to
randomization during crossovers. The direction of ${\bf J}$ is preserved in
the absence of random bombardment. {\it (b) Right panel}--
The applicability range for the Spitzer \& McGlynn (1979), Lazarian \& Draine (1997) and Lazarian \& Draine (1999ab) models.
The latter corresponds to the shaded area of "Flipping". The corresponding flipping rates were elaborated in Hoang \& Lazarian (2008b). $J_{max}$ parameterizes here the value of the pinwheel torques,
i.e. $J_{max}= t_{gas} dJ/dt$, where the derivative $dJ/dt$ arises from the pinwheel torque action.}
\label{range}
\end{figure}

P79  considered changes of the grain surface properties and noted that those
should stochastically change the direction (in body-coordinates) of the
systematic torques. Spitzer \& McGlynn (1979) developed a
theory of such {\it crossovers}. During a crossover, the grain slows down,
flips, and thereafter is accelerated again (see Fig.~\ref{range}a).

From the viewpoint of the grain-alignment theory, the important question is
whether or not a grain gets randomized during a crossover. If the value of the
angular momentum is small during the crossover, the grains are susceptible to
randomization arising from atomic bombardment. The original calculations in
Spitzer \& McGlynn (1979) obtained only marginal correlation between the values of the angular
momentum before and after a crossover, but their analysis disregarded thermal
fluctuations within the grain with temperature $T_{dust}$. According to the Fluctuation-Dissipation
Theorem these thermal fluctuation induce thermal wobbling with the frequency determined
by the internal relaxation rate (see Lazarian 1994). 
If the crossover happens over time shorter than the grain internal relaxation time, Lazarian \& Draine (1997) showed
that the Spitzer \& McGlynn (1979)
theory of crossovers should be modified to include the value of thermal angular momentum
$J_{d, \bot}\approx (2kT_{dust} I_{\|})^{1/2}$, where $I_{\|}$ is the maximal moment of inertial
of an oblate grain. What is the size of grains for which the effect gets important? When nuclear
relaxation is accounted for, this provides a size of $a_c\approx 10^{-4}$~cm. 
\subsection{Thermal trapping of grains}
\label{trapping}

What would happen to small grains? Lazarian \& Draine (1999a) predicted that
sufficiently small grains would flip fast and this will induce alternations of the directions  of pinwheel torques making grain
rotate thermally in spite of the presence of the uncompensated torques (see Fig.~\ref{relax}b).  
%

This way Lazarian \& Draine (1999a) explained why interstellar grains smaller than $10^{-6}$~cm are poorly aligned
via paramagnetic relaxation. A more elaborate study of the flipping phenomenon in Roberge \& Ford
(2000 preprint; see also Roberge 2004) supported this conclusion. Weingartner (2008), however, concluded
that the spontaneous thermal flipping does not happen if the internal relaxation
diffusion coefficient (Lazarian \& Roberge 1997) is obtained with a different integration constant. Does this
invalidate the phenomena of {\it thermal flipping} and {\it thermal trapping} predicted in Lazarian \& Draine (1999a)? Our study in
Hoang \& Lazarian (2008b) shows that the flipping and trapping do happen, if one takes into account additional
fluctuations associated, for instance, with the action of the uncompensated torques or gaseous bombardment.  

What does happen with the grains which do not flip, but smaller than $a_c$? For those grains the original
SM79 theory of crossovers is applicable and these grains are marginally aligned via Davis-Greenstein process, provided
that the pinwheel torques are short-lived, i.e. the time scale of their existence is much smaller than the time
of paramagnetic relaxation. Such grains rotate at high rate in accordance with the Purcell (1979) model.  As we discuss
further, the fact that pinwheel torques are not always suppressed by flipping may allow better alignment by radiative torques. 

Fig.~\ref{range}b defines the range over which the different model of crossovers are applicable. It is clear that for a sufficiently
large $J_{max}$ the flipping gets suppressed.

\section{Important Special Cases of Alignment}

\subsection{Radiative torque alignment of superparamagnetic grains}
\label{magnetic}

Superparamagnetic grains, i.e. grains with enhanced paramagnetic relaxation, 
 were invoked by Jones \& Spitzer (1967) within the model of paramagnetic alignment (see also Mathis 1986, Martin 1994, Goodman \& Whittet 1994, Roberge \& Lazarian 1999). 
What does happen when the dynamics of grains is determined by radiative torques? We see from Fig.~\ref{chi}b that 
for a substantial part of the parameter space grains are driven to the low-$J$ states, i.e. {\it subthermally}, which is in contrast to a widely spread belief 
(see Draine \& Weingartner 1996) that in the presence of radiative torques
most of the interstellar grains must rotate at  $T_{rot}\gg T_{gas}$.
\begin{figure}
\includegraphics[width=0.51\textwidth]{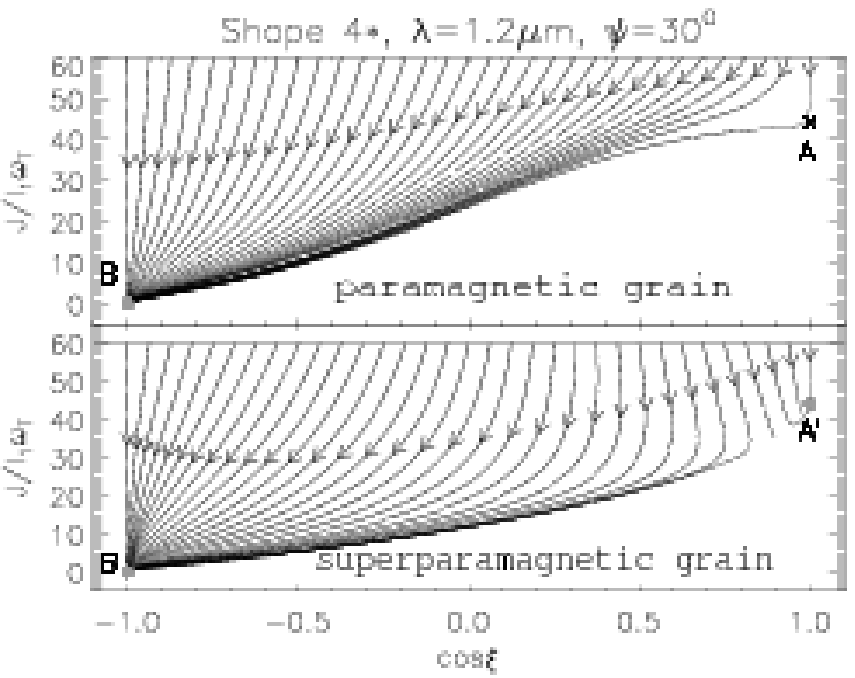}
\includegraphics[width=0.49\textwidth]{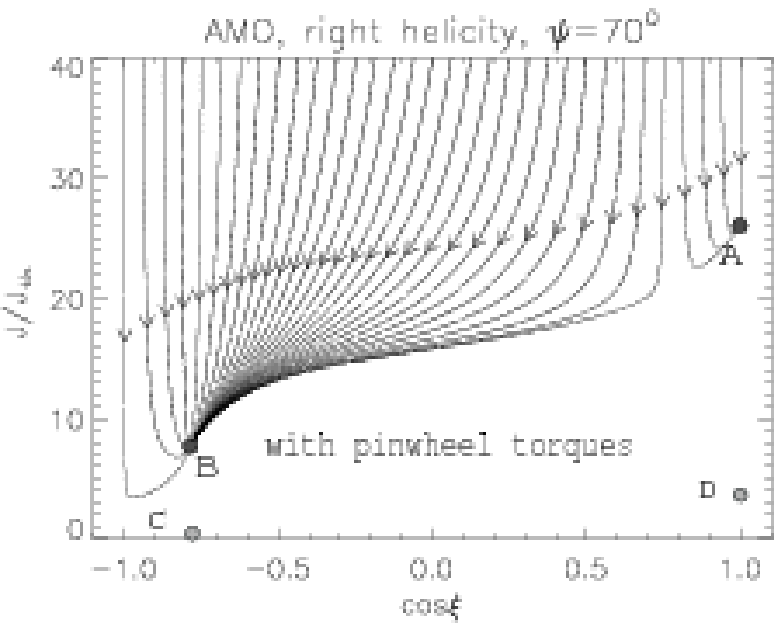}
\caption{\small {\it (a) Left Panel}-- A paramagnetic grain gets only a low-$J$ attractor point. For the same set of parameters a superparamagnetic grain gets also a high-$J$ attractor point.
The fact that most of the phase trajectories go in the direction of the low-$J$ attractor point illustrates the dominance of the radiative torques for the alignment even
in the case of the superparamagnetic grain. However, high-$J$ attractor points are more stable than the low-$J$ attractor points. As a result, all grains eventually end up
at the high-$J$ attractor point. From Lazarian \& Hoang (2007a). {\it Right Panel}-- Grain alignment by radiative torques in the presence
of  pinwheel torques. The shown case corresponds to the presence of both the low-$J$ and high-$J$ attractor points in the absence of pinwheel torques. In the case when only a low-$J$ attractor
point exists the strong pinwheel torques lift the low-$J$ attractor point enhancing the alignment. From Hoang \& Lazarian 2008b.}
\label{superparamagnetic}
\end{figure} 
The picture above, however, is different if grains paramagnetically dissipate the energy on the time scales shorter
than the gaseous damping time.  Lazarian \& Hoang (2008)  found that if grains have superparamagnetic inclusions, they {\it always} get  to a
high-$J$ attractor point. Fig.~\ref{superparamagnetic} shows that for superparamagnetic grains subject to a diffuse interstellar radiation field most grains still get
to the low attractor point, which reflects the fact that it is the radiative torques rather than paramagnetic ones that dominate the
alignment. As the high-$J$ attractor point is more stable compared to the low-$J$ attractor point
superparamagnetic grains get transfered by gaseous collisions from the low-$J$ to high-$J$ attractor points. Thus, superparamagnetic grains 
always rotate at high rate in the presence of radiative torques. One concludes that, rather
unexpectedly, intensive paramagnetic relaxation changes the rotational state of the grains, enabling them to
rotate {\it rapidly}. The alignment of grains at high-$J$ point is {\it perfect}.

\subsection{Radiative torque alignment in the presence of pinwheel torques}
\label{pinwheel}

Pinwheel torques were considered by Purcell (1979) in the context of paramagnetic alignment.
How do these torques also affect the radiative torque alignment? Hoang \& Lazarian (2008b) showed
that the sufficiently strong pinwheel torques can create new high-$J$ attractor points. Therefore for
sufficiently strong pinwheel torques, e.g. for torques arising from H$_2$ formation, one may observe the correlation
of higher degree of polarization with the atomic hydrogen content in the media, provided that H$_2$ torques as strong
as they considered in Purcell (1979) and the subsequent papers (see Spitzer \& McGlynn 1979, Lazarian 1995, Lazarian \& Draine 1997). 
Another implicit assumption for observing this correlation is that the grains are not superparamagnetic. For superparamagnetic grains
the alignment, as we discussed above, is perfect anyhow. 

The existing uncertainty is related to the magnitude of the pinwheel torques. If the correlation is observed, this would mean that, first of all,
grains are not superparamagnetic and, in addition, the pinwheel torques are sufficiently strong to overwhelm the radiative torques which drive
the grains to  low attractor points (see Figure~\ref{superparamagnetic}b).

\subsection{Alignment with negligible internal relaxation}

Our discussion in section~\ref{w1} indicates that for sufficiently large grains the internal relaxation over the time-scale of RAT alignment is negligible. Large grains, however, are present e.g. in accretion disks and comets. The polarization from accretion disks can provide one with an important insight into magnetic of the objects (see Cho \& Lazarian 2007), while understanding of the alignment for comet dust is important for explaining circular polarization observed (see below). Thus,
proper description of RAT alignment in the absence of internal relaxation is important. 

In the absence of internal relaxation can get aligned not only with long axes perpendicular to the angular momentum ${\bf J}$, but 
also with the longest axis parallel to ${\bf J}$, i.e. the axis of minimal moment of grain inertia parallel to ${\bf J}$. This complicates the analysis compared to the case of interstellar grains, for which the internal relaxation is very fast. The corresponding problem was addressed in Hoang \& Lazarian (2008c, henceforth HL08c). The results of the latter study are summarized in Table~1.
{\small
\begin{table}
\caption{Attractor points with and without internal relaxation for AMO. "L" denotes "long axes of the grain".}
\begin{tabular}{llcll} \hline\hline\\
\multicolumn{1}{c}{\it Without relaxation (HL08c)} & & \multicolumn{1}{c}{\it With relaxation (LH07a)}\\[1mm]
\hline\\
{{\bf High-J }}& {{\bf Low-J }} &{{\bf High-J}}&{{\bf Low-J}}\\[1mm]
&&&&\\[1mm]
{$\bJ\|\bB$}& {$\bJ$ $\|$ or at} &{$\bJ\| \bB$}&{$\bJ$  $\|$ or at}\\[1mm]
&{angle with $\bB$}&&{angle with $\bB$}\\[1mm]
{L $\perp \bB$}& {L $\perp$ or $\|$ $\bJ$} &{L $\perp \bB$}&{L $\perp \bB$}\\[1mm]\\[1mm]
\hline\hline\\
\end{tabular}
\end{table}
}
The study of HL08c is suggestive that grains do preferentially get aligned with long axes perpendicular to magnetic field even without internal relaxation. Indeed,
this is consistent with the finding there that such alignment happens for the high-$J$ attractor points of AMO. If this is the case, when the low-$J$ and high-$J$ attractor points coexist the steady-state alignment will happen only with high-$J$ attractor point (see discussion in \S\ref{magnetic}).  The wrong alignment happens only with low-$J$ attractor points and it is suggestive that in the presence of gaseous bombardment the grains may still spend more time in the vicinity of the high-$J$ repellor point, as was shown in numerical simulations by Hoang \& Lazarian (2008a). The conclusions obtained with AMO are consistent with a limited parameter study obtained in HL08c for an irregular grain. However, it is clear that more extensive studies of the RAT alignment in the absence of internal relaxation are required.

\section{Application of Grain Alignment Theory: Examples}

\subsection{Alignment in molecular clouds and diffuse/dense cloud interface}

\begin{figure}
\includegraphics[width=0.5\textwidth]{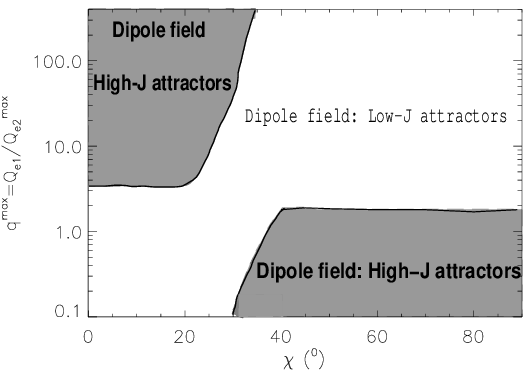}
\includegraphics[width=0.5\textwidth]{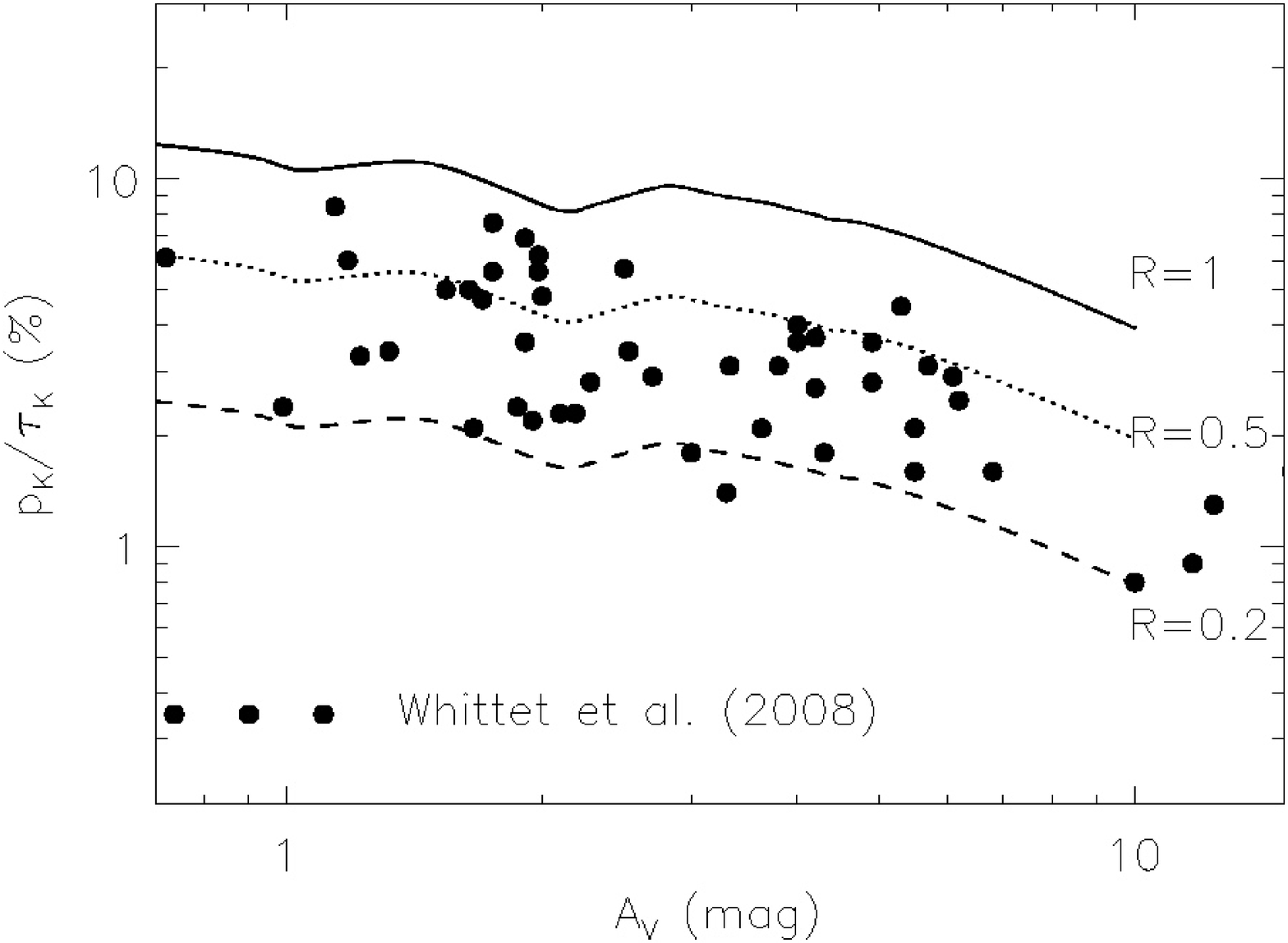}
\caption{\small {\it (a) Left Panel}-- The parameter space for the existence of only low-$J$ and both low-$J$ and high-$J$ attractor points for the case of grains subject to external dipole radiation. Angle $\xi$ is between the dipole axis and the direction of external magnetic field.  A comparison with
Figure~\ref{chi}b reveals that the parameter space is extended compared to the case of radiation from a single source ("beam"). Radiation field in clouds and disks can be decomposed into dipole and quadrupole components. For the latter the parameter space for the existence of high-$J$ attractor points and thus for perfect alignment, has also been defined. From Hoang \& Lazarian (2008b). {\it Right Panel}-- Theoretical predictions for grain alignment by RATs in the presence of turbulence compared with Whittet et al. (2008) data. $R$ is the Rayleigh reduction factor which characterizes the alignment. It is clear that the alignment of 20\% is too weak to explain the data.  }
\label{cloud}
\end{figure}

Aligned grains provide the best way to trace magnetic fields in molecular clouds and getting insight into the role of magnetic field in star formation (see Hildebrand et al. 2000). In dense molecular clouds one might expect all grain alignment mechanisms to be suppressed as conditions there approach those of
thermal equilibrium (see discussion in Lazarian et al. 1997). This conclusion, however, contradicts to observations in Ward-Thompson et al. (2000), which revealed aligned grains dense in pre-stellar cores with high (up to $A_v=15$) optical depth. Cho \& Lazarian (2005) noticed that the efficiency of grain alignment is a function of the grain size (see Figure~\ref{wavelength}b), and this enables even highly attenuated reddened interstellar light to aligne large grains that are known to exist in dark cores. More detailed calculations in Bethell et al. 2007 supported this explanation. However, even the latter paper which considered the inhomogeneity of density distribution for the purposes of the radiation transfer, used only crude estimates of grain alignment efficiencies. Future studies should capitalize on the recent progress of grain alignment theory. For instance, it is advisable to take into account that the attenuated light in molecular clouds can be decomposed into the dipole and quadrupole components and take into account the theoretical predictions for the alignment by these components obtained in HL08c. For example, comparing Figures~\ref{chi}b and \ref{cloud}a one can see that the alignment by the dipole component corresponds to a larger, compared to a beam, parameter space with high-$J$ attractor points. 
Modeling of variations of the dependence of polarization degree with wavelengths, the dependence termed by Hildebrand et al. (2000) "polarization spectrum" is another important test of the grain alignment. Calculations of the polarization spectrum in the absence of embedded stars are presented in Lazarian (2007), while those in the presence of stars are made in Chepurnov \& Lazarian (2009). The latter exhibit better correspondence with the observations.

In terms of polarization in diffuse media, the RAT alignment reproduces well the empirical Serkowski (1973) law (see Lazarian 2007). Moreover, RATs can explain the change of the polarization degree observed at the interface of the diffuse media and a molecular cloud. The latter variations were first reported in
Whittet et al. (2001) and explained as a consequence of the RAT-induced alignment in Lazarian (2003). A more extended data set was analyzed in Whittet et al. (2008), where a quantitative correspondence between the RAT predictions and the observations was established. A fit to Whittet et al. (2008) data, which, apart from the RATs efficiencies, takes into account the effects of magnetic field turbulence is shown in Figure~\ref{cloud}b.  

\subsection{Alignment of grains and polarization of Zodiacal  light}

Circular polarization of Zodiacal light was reported in Wolstencroft \& Kemp (1973). The authors explained this emission as arising from scattering of unpolarized Solar radiation by aligned grains.  Later Dolginov \& Mytrophanov (1976) briefly discussed in the framework of existed at that time paradigms possible alignment processes of Zodiacal dust. One way or another, the alignment of Zodiacal dust has been mostly ignored, which now presents a problem for studies of polarized CMB foreground emission.

In Figure~\ref{circular}a we present our calculations of the expected circular polarization $p_{V}/p_{V}^{max}$ as a function of elongation angle $\epsilon$ between the direction from the Earth to the Sun and the line of sight, for an intentionally simplified model of the interplanetary magnetic field. The direction of magnetic field is assumed to incline by an angle of 10 degree with the eliptic plane, and the component parallel to the eliptic plane makes a constant angle with respect to the all lines of sight. It can be seen that $p_{V}/p_{V}^{max}$ changes its sign at $\epsilon\sim 90, 180, 220, 270$ and $320$ degree. By comparing $p_{V}/p_{V}^{max}$ from calculations with observation data, one can infer the magnetic field geometry in the interplanetary medium.

Could there be directions with no circular polarization of Zodiacal light and a substantial polarized contribution from the Zodiacal dust to the microwave emission? The answer to this question is positive. Indeed, grain alignment theory predicts efficient grain alignment in the Solar neiborhood. If magnetic field are exactly in the plane of Earth's ecliptic, then the circular polarization is zero. Indeed, the circular polarization is proportional to scalar product of magnetic field direction vector ${\bf B}/|B|$ and the vector product of ${\bf n}_E\times {\bf n}_S$, where ${\bf n}_E$ gives the direction from the scattering volume to the Earth and ${\bf n}_S$ provides the direction from the Sun to the scattering volume. When all three vectors are in the same plane, the result is zero, in spite of the fact that the grains are aligned.

The measurements in Wolstencroft \& Kemp (1973) show that in most cases magnetic field is at an angle to the Earth ecliptic, which provides a means of studying the variations of the magnetic field both in space and time. The long range variations of magnetic field should provide input on the injection scale of Solar wind turbulence, which is, by itself is an important problem. Needless to say, that the modeling of the polarization from Zodiacal light is essential for successful separating of this contribution from the polarized CMB.  

\subsection{Alignment of grains in comet atmosphere}

The ``anomalies'' of polarization from comets\footnote{When light is scattered
by the randomly oriented particles with sizes much
less than the wavelength, the scattered light is
 polarized perpendicular to the scattering plane, which is the
plane passing through the Sun, the comet and the observer.
Linear polarization from comets has been long known to exhibit
polarization that is not perpendicular to the scattering plane.}
(see  Martel 1960, Beskrovnaja et al 1987, Ganesh et al 1998) as well as circular
polarization from comets (Metz \& Haefner 1987, Dollfus \& Suchail 1987, Morozhenko et al 1987)
are indicative of grain alignment.
\begin{figure}
\includegraphics[width=0.45\textwidth]{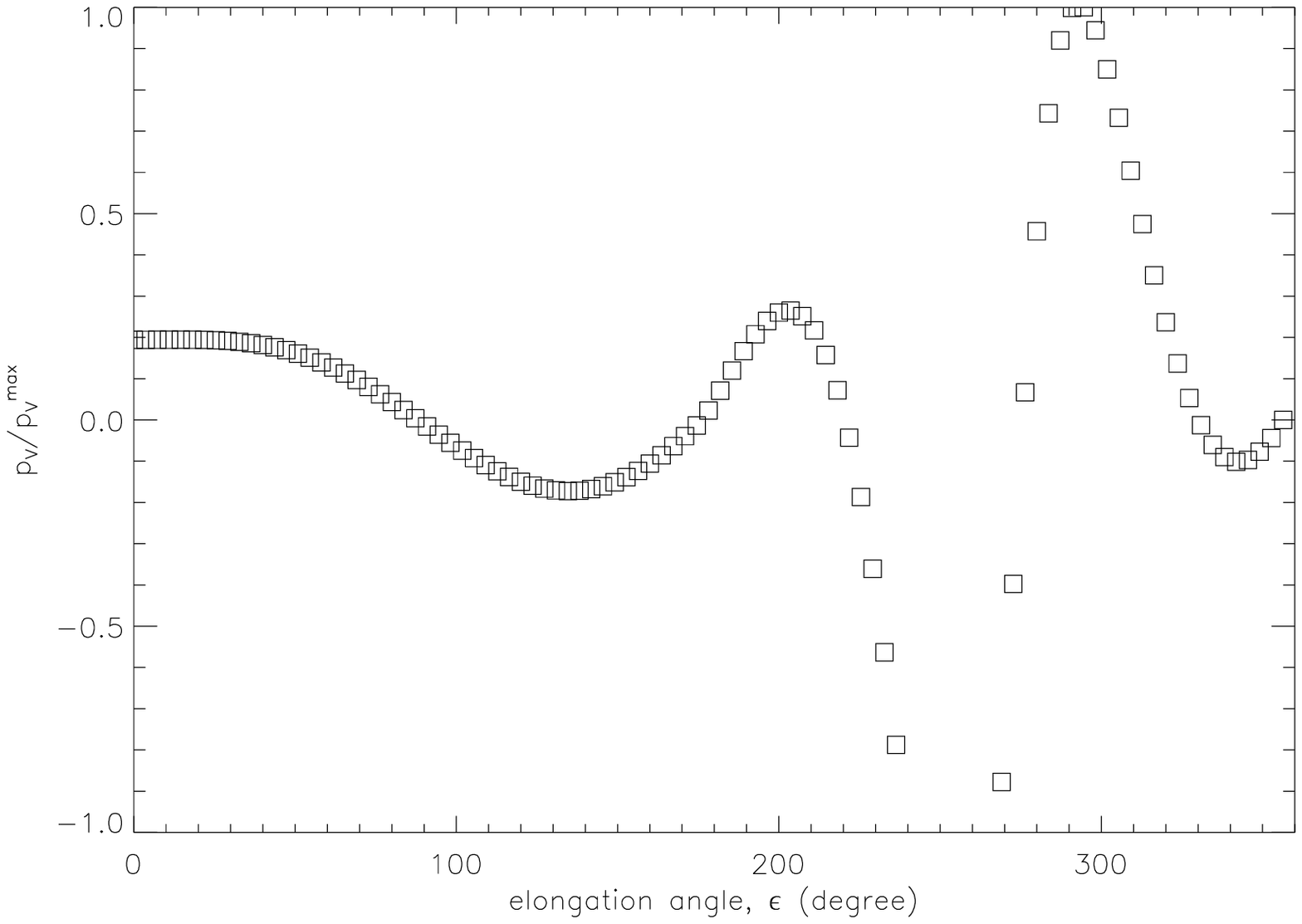}
\includegraphics[width=0.45\textwidth]{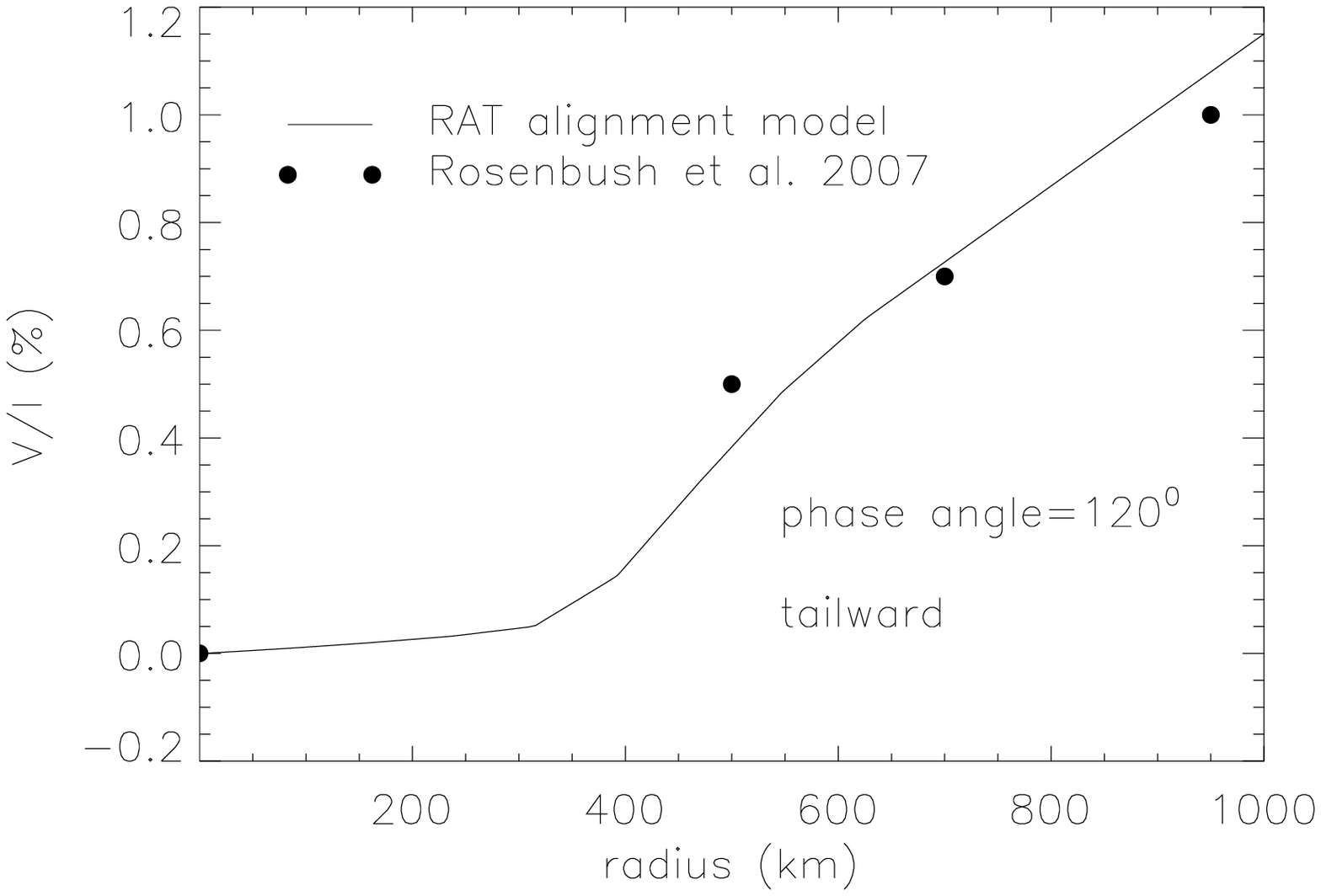}
\caption{\small {\it Circular polarization. (a) Left Panel}--  Normalized polarization $p/p_{max}$ of the Zodiacal light for the model with magnetic field  inclined by an angle of 10 degree with the eliptic plane, and the component parallel to the eliptic plane makes a constant angle with respect to all lines of sight. More
realistic modeling should use a spiral structure of magnetic field.  {\it Right Panel}-- Comparison of circular polarization for the model with observations in Rosenbush et al. (2007). The distance is measured from the nucleus of the comet.}
\label{circular}
\end{figure}     
However, conclusive arguments in favor of grain alignment 
were produced for the Levi (1990 20) comet through direct
measurements of starlight polarization, as the starlight was passing
through comet coma (Rosenbush et al 1994). The data conclusively
proved the existence of aligned grains in comets.

Note, that the issue of circular polarization was controversial in the past.
When both left and right handed polarization is present in different
parts of coma the average over entire coma may get the 
polarization degree close to zero.
This probably explains why earlier researchers were unsuccessful attempting
to measure circular polarization while using large
apertures. Recent measurements by Rosenbush et al. (1999), Manset et al.
(2000) of circular polarization from Hale-Bopp Comet support the notion
that circular polarization is a rule rather than an exception.

A more recent paper by Rosenbush et al. (2006) reports circular polarization
from a comet C/1999 S4(LINEAR). The data indicates that the polarization arises
from aligned grains.  
Figure~\ref{circular}b shows our predictions for the circular polarization versus observations. Our predictions are
obtained in a simple model where grains have non-zero electric dipole moments (see Draine \& Lazarian 1998) and precess in the
radial electric field of the comet (see Serezhkin 2000). The grains are being aligned by RATs.
The encouraging correspondence between the observations and theory call for detailed modeling of circular polarization from comets.

\section{Discussion}  

We would like to stress two new recent developments related to grain alignment theory. First of all, the theory has become quantitative and, for the
first time, the theory is able to account for the existing observational data. Another important development is that the grain alignment is no longer a
concern of interstellar experts only. Examples in this review as well as in Lazarian (2007) convincingly show that one should account for aligned grains dealing with different environments, e.g. accretion disks, circumstellar regions, interplanetary space and comets. Understanding of the polarization from dust in diffuse media
is also a burning issue for the CMB polarization research. 

Our review above is devoted to the recent developments in the theory of RAT alignment. However, we know the limitations of the RAT mechanism. For instance,
RATs are inefficient in aligning grains for which $a_{eff}\ll \lambda$ (see Figure~\ref{wavelength}b). Therefore for sufficiently small grains other mechanisms should be important. For instance, Lazarian \& Draine (2000) introduced a new process, which they termed "resonance paramagnetic relaxation" and showed that this process can be important for the alignment of very small grains or PAHs, which are responsible for the galactic anomalous microwave emission (Draine \& Lazarian 1998). We believe that there are other circumstances when other alignment processes take over. For instance, the mechanical alignment of helical grains introduced by Lazarian \& Hoang (2007b) may compete with the RAT alignment in particular circumstances. Future research should
identify, on the basis of comparing theoretical predictions and observations, these circumstances. 

As it is not possible to cover the subject of grain alignment within a short review, we address our reader to earlier reviews, where particular facets of the grain alignment problem are presented in more detail. The most detailed review to this moment is Lazarian (2007), which, however, in terms of the alignment theory, does not cover the most recent developments. Thus Lazarian (2007) is complementary to the present review. Reviews written earlier than 2007 are mostly outdated in terms of the RAT alignment. However, Lazarian (2003) review may be recommended for those who are interested in the history of grain alignment ideas. The connection between grain alignment and CMB polarization studies is discussed in Lazarian (2008) review.


\acknowledgements 
Our research was supported by the NSF Center for Magnetic Self-Organization in Astrophysical and Laboratory Plasmas and the NSF grant 0507164.


\end{document}